\documentclass[12pt,aps,prd,showpacs,amsmath,amssymb]{revtex4}
\input epsf
\begin{document}
\title{Higher order corrections to the light-cone distribution amplitudes of the sigma baryon}
\author{Yong-Lu Liu$^1$, Chun-Yu Cui$^2$, and Ming-Qiu Huang$^1$}
\affiliation{$^1$ College of Science, National University of Defense Technology, Hunan 410073, China}
\affiliation{$^2$ Department of Physics, School of Biomedical Engineering, Third Military Medical University, Chongqing 400038, China}
\date{\today}
\begin{abstract}
The light-cone distribution amplitudes (LCDAs) of the $\Sigma^\pm$ baryons up to twist-$6$ are investigated on the basis of the QCD conformal partial wave expansion approach. The calculations are carried out to the next-to-leading
order of conformal spin accuracy. The nonperturbative parameters relevant to the LCDAs are determined in the framework of the QCD sum rule method. The explicit expressions of the LCDAs are given as the main results.
\end{abstract}
 \maketitle

\section{Introduction}\label{sec1}

Signals confirmed by ATLAS and CMS\cite{HiggsCMS,HiggsATLAS} showed that the Higgs boson\cite{Higgs} in the standard model (SM) have been found and the SM is most likely to be a precise theory at the present energy scale. New physics beyond the SM at higher energy scale is mostly concerned nowadays and in the near future. However, many difficulties are still alive in practical analysis of hadron physics involving nonperturbative QCD effect when we study hadronic phenomena at low energy, or $\Lambda_{QCD}$ scale. A typical method to solve the nonperturbative difficulties in QCD is factorization, in which the nonperturbative part is included into the wave function, such as the parton distribution functions for inclusive processes, fragmentation distribution functions for the hadronization, and the distribution amplitudes for exclusive processes. Specifically, in theoretical investigations of the hard exclusive processes \cite{exclusive,exclusive2} and hadronic physics with the QCD light-cone sum rule method \cite{lcsr1,lcsr2,lcsr3}, the light-cone distribution amplitudes (LCDAs) are fundamental ingredients to be studied. Furthermore, when searching for new physics beyond the SM, it is an important way to study flavor physics, in which some processes that are sensitive to the new physics can be measured more precisely nowadays than any time before. All of these require detailed information of the internal structure and the dynamical properties of the hadron, which are dominated by the nonperturbative QCD characters.

In the past decades, many efforts have been made in the descriptions of mesons\cite{mesondas} and the nucleon\cite{Chernyak,Braun1,Lenz,dalattice,Pasquini,overviewofDAs}, whereas, theoretical studies of a large number of the hadron physics phenomena require us to know LCDAs of many other hadrons such as the octet baryons, the decuplet baryons, and some excited hadron states that are difficult to be determined experimentally at present. We have examined the LCDAs of the strange octet baryons in the previous work\cite{DAs} in the conformal spin expansion method\cite{Braun2,Balitsky,Lenz}. Our calculation concerns LCDAs to twist-$6$ to the accuracy of the leading order of conformal spin expansion. The obtained parameters are also used to analyze some hadronic physics processes as applications\cite{LYL,LYL2,LYL3,Aliev}. However, some of the investigations\cite{LYL,LYL3} have implied that corrections from the higher order conformal spin contributions may affect the results to some extent.

In the point view of applications, an important effect to the LCDAs is the correction of the higher twist distribution functions. The higher order twist contributions to LCDAs have several origins, among which the main one comes from ``bad'' components in the wave function and in
particular of components with ``wrong'' spin projection for the case of baryons \cite{Braun1,DAs}. Compatible with the previous works, we focus on higher order twist contributions from bad components in the decomposition of the Lorentz structure in this paper. One of the general descriptions of LCDAs is based on the conformal symmetry of the massless QCD Lagrangian dominated on the light cone. The conformal partial wave expansion of the LCDAs can be carried out safely in the limit of the $SU(3)$-flavor symmetry approach. However, when terms connected with the $s$-quark mass are considered, the $SU(3)$-flavor breaking effects need to be included. In the present work, effects from the $SU(3)$-flavor symmetry breaking are considered as the corrections, which originate from two sources: isospin symmetry breaking and corrections to the nonperturbative parameters.

It is known that the leading order contribution with the conformal spin expansion approach comes from the properties of the matrix elements of the local three-quark operator between the vacuum and the baryon state. Thus, it is natural that higher order corrections should be related to the expansion of the matrix elements of the nonlocal three-quark operator at the zero point. However, we still need to estimate how much the contributions from four-particle effects will do on the result. Fortunately we have known that for processes whose dominant contribution is from the light cone the four-particle contributions can be safely omitted in the lower leading order. Thus, in the following analysis we only consider contributions from three-quark operator matrix element, whose higher moment is calculated with QCD sum rules\cite{SVZ}.

As applications, the light-cone QCD sum rule method has been used to examine processes related to the strange octet baryons and give instructive estimates \cite{Aliev,Wang}. In the previous works, we have analyzed some physical processes related to the final states about the $\Sigma$ baryon. The results are compatible with the experiments and(or) the other theoretical predications\cite{LYL2}. Nevertheless, there are still some processes which are not well described \cite{LYL,LYL3}. We wish the higher order corrections from the higher conformal spin may give us more accurate estimates.

The rest of the paper is organized as follows. Section \ref{sec:def} is devoted to present the definitions of the higher order moment of the three-quark operators related to corrections of LCDAs from the higher conformal spin expansion. In Sec. \ref{sec:conexp}, the conformal partial wave expansion of the LCDAs is carried out by use of the conformal symmetry of the massless QCD Lagrangian. The nonperturbative parameters connected with the LCDAs are determined in Sec. \ref{sec:sumrule} with the QCD sum rule method. Finally, we give the explicit expressions of the $\Sigma$ baryon LCDAs in Sec. \ref{sec:result}. A summary is given is Sec. \ref{sec:summary}. The equations of motion which are used to reduce the number of the free parameters are presented in Appendix \ref{app:1} for the completeness of this paper. The sum rule of one coupling constant $V_1^s$ is analyzed in Appendix \ref{app:2} as an example to elucidate the principal process of this method, and the other sum rules can be carried out in the same way.

\section{Higher conformal expansion of the light-cone distribution amplitudes of $\Sigma$}\label{sec:def}

\subsection{General definition}
Matrix elements of the quark-quark or quark-gluon-quark field operator between vacuum or hadron states are the great important ingredients in analysis of processes in quantum field theory. Light-cone distribution amplitudes of the $\Sigma$  baryon are defined by the general Lorentz expansion of the matrix element of the nonlocal three-quark-operator between the vacuum and the baryon state
\begin{equation}
\langle{0} |\epsilon^{ijk} q_\alpha^i(a_1 z) q_\beta^j(a_2 z) s_\gamma^k(a_3 z) |{\Sigma(P)}\rangle \,,\label{matele}
\end{equation}
where $q$ represents $u$ or $d$ quark, which correspond to $\Sigma^+$ or $\Sigma^-$ baryon, respectively. The indices $\alpha, \beta, \gamma$ refer to Lorentz indices and $i, j, k$ represent color ones. It is noticed that to make the matrix element above gauge invariant, the gauge factor $[x,y]=P\mbox{exp}[ig_s\int_0^1dt(x-y)_\mu A^\mu(tx+(1-t)y)]$ need to be inserted, whereas when fixed-point gauge $(x-y)^\mu A_\mu(x-y)=0$ is  adopted, this factor is equal to unity. Thus in this paper we do not show them explicitly.

Taking into account the Lorentz covariance, spin and parity properties of the baryons, the matrix element (\ref{matele}) is generally decomposed as
\begin{equation}
4 \langle{0} |\epsilon^{ijk} q_\alpha^i(a_1 z) q_\beta^j(a_2 z) s_\gamma^k(a_3 z) |{\Sigma(P)}\rangle = \sum\limits_{i}\mathcal{F}_i\,\Gamma_{1i}^{\alpha\beta}\Big
(\Gamma_{2i}\Sigma\Big )_\gamma \,,\label{da-def}
\end{equation}
where $\Sigma_\gamma$ is the spinor of the baryon with the quantum number $I(J^P)=1(\frac{1}{2}^+)$ ($I$ is the isospin, $J$ is the total angular momentum, and
$P$ is the parity), $\Gamma_{1(2)i}$ are certain Dirac structures over which the sum is carried out, and $\mathcal{F}_i=\mathcal{S}_i, \mathcal {P}_i, \mathcal
{A}_i, \mathcal{V}_i,\mathcal{T}_i$ are the independent distribution amplitudes which are functions of the scalar product $P\cdot z$\cite{DAs}. It is also noticed that $z$ and $p$ are the two light-cone vectors which satisfy $z^2=0$ and $p^2=0$.

Functions defined above do not have definite twist. In order to classify the LCDAs according to the definite twist, we redefine the wave functions $\mathcal{F}_i$ in the infinite momentum frame as:
\begin{equation}
4\langle {0}| \epsilon^{ijk} {s_1}_\alpha^i(a_1 z) {s_2}_\beta^j(a_2 z) q_\gamma^k(a_3 z) |{\Sigma(P)}\rangle =\sum\limits_{i}F_i\,\Gamma_{1i}'^{\alpha\beta}\Big
(\Gamma_{2i}'\Sigma\Big )_\gamma\,.\label{da-deftwist}
\end{equation}
 A naive calculation shows that the invariant functions $\mathcal S_i,\mathcal P_i,\mathcal V_i,\mathcal A_i,\mathcal T_i$ can be expressed in terms of the LCDAs $F_i=S_i,P_i,V_i,A_i,T_i$. The two sets of definitions have the following relations:
\begin{eqnarray}
&{\cal S}_1 = S_1\,, & 2p\cdot z\,{\cal S}_2 = S_1-S_2\,, \nonumber\\
&{\cal P}_1 = P_1\,, & 2p\cdot z\,{\cal P}_2 = P_2-P_1\,,\nonumber\\
&\mathcal V_1 = V_1\,, & 2 p\cdot z\mathcal V_2 = V_1 - V_2 - V_3\,,\nonumber \\
&2 \mathcal V_3 = V_3\,, & 4 p\cdot z\mathcal V_4 = - 2 V_1 + V_3 + V_4 + 2 V_5\,,\nonumber \\
&4 p\cdot z \mathcal V_5 = V_4 - V_3\,, & (2 p\cdot z )^2\mathcal V_6 = - V_1 + V_2 + V_3 + V_4 + V_5 - V_6
\end{eqnarray}
for scalar, pseudoscalar, and vector structure, and
\begin{eqnarray}
&\mathcal A_1 = A_1\,, & 2 p\cdot z\mathcal A_2 = - A_1 + A_2 - A_3\,, \nonumber\\
&2 \mathcal A_3 = A_3\,, & 4 p\cdot z\mathcal A_4 = - 2 A_1 - A_3 - A_4 + 2 A_5\,, \nonumber\\
&4 p\cdot z \mathcal A_5 = A_3 - A_4\,, & (2 p\cdot z )^2\mathcal A_6 = A_1 - A_2 + A_3 + A_4 - A_5 + A_6
\end{eqnarray}
for axial-vector structure, and
\begin{eqnarray}
&\mathcal T_1 = T_1\,, & 2 p\cdot z\mathcal T_2 = T_1 + T_2 - 2 T_3\,, \nonumber\\
&2 \mathcal T_3 = T_7\,,& 2 p\cdot z\mathcal T_4 = T_1 - T_2 - 2 T_7\,, \nonumber\\
&2 p\cdot z \mathcal T_5 = - T_1 + T_5 + 2 T_8\,, &
(2 p\cdot z)^2\mathcal T_6 = 2 T_2 - 2 T_3 - 2 T_4 + 2 T_5 + 2 T_7 + 2 T_8\,, \nonumber\\
&4 p\cdot z \mathcal T_7 = T_7 - T_8\,, &(2 p\cdot z)^2\mathcal T_8 = -T_1 + T_2 + T_5 - T_6 + 2 T_7 + 2 T_8
\end{eqnarray}
for tensor structure.

The classifications of the LCDAs $F_i$ with a definite twist are listed in Table \ref{tabDA-def}, where we take $\Sigma^+$ as an example. The explicit expressions of the definition can be found in Refs. \cite{Braun1,DAs}. Each distribution amplitude $F_i$ can be represented as
\begin{equation}
F(a_ip\cdot z)=\int \mathcal Dxe^{-ipz\sum_ix_ia_i}F(x_i),
\end{equation}
where the dimensionless variables $x_i$, which satisfy the relations $0<x_i<1$ and $\sum_ix_i=1$, correspond to the longitudinal momentum fractions along the light cone carried by the quarks inside the baryon. The integration measure is defined as
\begin{equation}
\int \mathcal Dx=\int_0^1dx_1dx_2dx_3\delta(x_1+x_2+x_3-1).
\end{equation}
\begin{table}
\renewcommand{\arraystretch}{1.1}
\caption{Independent baryon distribution amplitudes in the chiral expansion.}
\begin{center}
\begin{tabular}{|l|l|l|l|}
\hline& Lorentz-structure & Light-cone projection & Nomenclature
\\ \hline
Twist-3 &  $ \left(C\!\not\!{z}\right) \otimes \!\not\!{z} $ & $u^+_\uparrow u^+_\downarrow s^+_\uparrow$ & $\Phi_3(x_i) = \left[V_1-A_1\right](x_i)$ \\ \hline&  $
\left(C i \sigma_{\perp z} \right) \otimes \gamma^\perp \!\not\!{z} $ & $u^+_\uparrow u^+_\uparrow s^+_\downarrow$ & $T_1(x_i)$ \\  \hline Twist-4 &  $
\left(C\!\not\!{z}\right) \otimes \!\not\!{p} $ & $u^+_\uparrow u^+_\downarrow s^-_\uparrow$ & $\Phi_4(x_i) = \left[V_2-A_2\right](x_i)$
\\ \hline
& $ \left(C\!\!\not\!{z}\gamma_\perp\!\!\not\!{p}\,\right)
      \otimes \gamma^\perp\!\!\not\!{z} $
& $ u^+_\uparrow u^-_\downarrow s^+_\downarrow$ & $\Psi_4(x_i) = \left[V_3-A_3\right](x_i)$ \\ \hline & $ \left(C \!\not\!{p}\!\not\!{z}\right)  \otimes
\!\not\!{z}$ & $u^-_\uparrow u^+_\uparrow  s^+_\uparrow$ & $\Xi_4(x_i) = \left[T_3-  T_7 + S_1 + P_1\right](x_i)$ \\ \hline & $ \left(C
\!\not\!{p}\!\not\!{z}\right) \otimes \!\not\!{z}$ & $u^-_\downarrow u^+_\downarrow  s^+_\uparrow$ & $\Xi_4'(x_i) = \left[T_3 + T_7 + S_1 - P_1\right](x_i)$\\
\hline& $ \left(C i \sigma_{\perp z} \right) \otimes \gamma^\perp \!\not\!{p} $ & $u^+_\downarrow u^+_\downarrow s^-_\downarrow$ & $T_2(x_i)$
\\ \hline
Twist-5 &  $ \left(C\!\not\!{p}\right) \otimes \!\not\!{z} $ & $u^-_\uparrow u^-_\downarrow s^+_\uparrow$ & $\Phi_5(x_i) = \left[V_5-A_5\right](x_i)$
\\ \hline
& $ \left(C\!\!\not\!{p}\gamma_\perp\!\!\not\!{z}\,\right) \otimes \gamma^\perp\!\!\not\!{p} $ & $ u^-_\uparrow u^+_\downarrow
s^-_\downarrow $ & $\Psi_5(x_i) = \left[V_4-A_4\right](x_i)$ \\
\hline & $ \left(C \!\not\!{z}\!\not\!{p}\right)  \otimes \!\not\!{p}$ & $u^+_\uparrow u^-_\uparrow  s^-_\uparrow$ &
$\Xi_5(x_i) = \left[-T_4- T_8 + S_2 + P_2\right](x_i)$ \\
\hline & $ \left(C \!\not\!{z}\!\not\!{p}\right)  \otimes \!\not\!{p}$ & $u^+_\downarrow u^-_\downarrow  s^-_\uparrow$ & $\Xi_5'(x_i) = \left[S_2 - P_2-T_4+
T_8\right](x_i)$ \\ \hline&  $ \left(C i \sigma_{\perp p} \right) \otimes \gamma^\perp \!\not\!{z} $ & $u^-_\downarrow u^-_\downarrow s^+_\downarrow$ & $T_5(x_i)$
\\ \hline
Twist-6 &  $ \left(C\!\not\!{p}\right) \otimes \!\not\!{p} $ & $u^-_\uparrow u^-_\downarrow s^-_\uparrow$ & $\Phi_6(x_i) = \left[V_6-A_6\right](x_i)$ \\ \hline&  $
\left(C i \sigma_{\perp p} \right) \otimes \gamma^\perp \!\not\!{p} $ & $u^-_\uparrow u^-_\uparrow s^-_\downarrow$ & $T_6(x_i)$ \\ \hline
\end{tabular}
\end{center} \label{tabDA-def}
\end{table}

There exist some symmetry properties of the LCDAs from the identity of the two $u/d$ quarks in the $\Sigma$ baryon, which is useful to reduce the number of the independent functions. Taking into account the Lorentz decomposition of the $\gamma$-matrix structure, it is easy to see that the vector and tensor LCDAs are symmetric, whereas the scalar, pseudoscalar and axial-vector structures are antisymmetric under the interchange of the two $u/d$ quarks:
\begin{eqnarray}
V_i(1,2,3)&=&\;\ V_i(2,1,3),\hspace{2.8cm} T_i(1,2,3)=\;\ T_i(2,1,3),\nonumber\\
S_i(1,2,3)&=&-S_i(2,1,3),\hspace{2.7cm} P_i(1,2,3)=-P(2,1,3),\nonumber\\
A_i(1,2,3)&=&-A(2,1,3).
\end{eqnarray}
The similar relationships hold for the ``calligraphic'' structures in Eq. (\ref{da-def}).

In order to expand the LCDAs by the conformal partial waves, we rewrite the LCDAs in terms of quark fields with definite chirality
$q^{\uparrow({\downarrow})}=\frac{1}{2}(1\pm\gamma_5)q$. Taking $\Sigma^+$ as an example, the classification of the LCDAs in this presentation can be interpreted transparently: projection on the
state with the two $u$-quarks antiparallel, i.e. $u^\uparrow u^\downarrow$, singles out vector and axial vector structures, while parallel ones, i.e. $u^\uparrow
u^\uparrow$ and $u^\downarrow u^\downarrow$, correspond to scalar, pseudoscalar and tensor structures. The explicit expressions of the LCDAs by chiral-field
representations are presented in Table \ref{tabDA-def} as an example for $\Sigma^+$. The counterparts of $\Sigma^-$ can be easily obtained under the exchange
$u \rightarrow d$.

Note that in the case of the nucleon, the isospin symmetry can be used to reduce the number of the independent LCDAs to eight\cite{Braun1}. However, there are no similar isospin symmetric relationships existing when the $\Sigma$ baryon is considered. Therefore, we need altogether $14$ chiral field representations to express all the LCDAs.

\subsection{Conformal expansion}\label{sec:conexp}

In this subsection we give the explicit form of the LCDAs with the aid of the conformal partial wave expansion approach. The main idea of this method is based on the conformal symmetry of the massless QCD Lagrangian. In this approach the longitudinal degrees of freedom can be separated from transverse ones. On the one hand, the properties of transverse coordinates are described by the renormalization scale that is determined by the renormalization group equation. On the other hand, the longitudinal momentum fractions that are living on the light cone are governed by a set of orthogonal polynomials, which form an irreducible representation of the collinear subgroup $SL(2,R)$ of the conformal group.

The algebra of the collinear subgroup $SL(2,R)$ is determined by the following four generators:
\begin{equation}
{\bf L}_+=-i{\bf P}_+,\; {\bf L}_-=\frac{i}{2}{\bf K}_-,\; {\bf L}_0=-\frac{i}{2}({\bf D}-{\bf M}_{-+}),\; {\bf E}=i({\bf D}+{\bf M}_{-+}),
\end{equation}
where ${\bf P}_\mu$, ${\bf K}_\mu$, $\bf D$, and ${\bf M}_{\mu\nu}$ correspond to the translation, special conformal transformation, dilation and Lorentz generators, respectively. The notations are used for a vector $A$: $A_+=A_\mu z^\mu$ and $A_-=A_\mu p^\mu/p\cdot z$. Let ${\bf L}^2={\bf L}_0^2-{\bf L}_0+{\bf L}_+{\bf L}_-$, then a given distribution amplitude with a definite twist can be expanded by the conformal partial wave functions that are the eigenstates of ${\bf L}^2$ and ${\bf L}_0$.

For the three-quark state, the distribution amplitude with the lowest conformal spin $j_{min}=j_1+j_2+j_3$ is \cite{Braun2,Balitsky}
\begin{equation}
\Phi_{as}(x_1,x_2,x_3)=\frac{\Gamma[2j_1+2j_2+2j_3]}{\Gamma[2j_1]\Gamma[2j_2]\Gamma[2j_3]}{x_1}^{2j_1-1}{x_2}^{2j_2-1}{x_3}^{2j_3-1},\label{as-dis}
\end{equation}
where $j_i$ represents the conformal spin of the quark field that is defined as half of the canonical dimension plus its spin $j=(l+s)/2$. Contributions with higher conformal spin $j=j_{min}+n$ ($n=1,2,...$) are given by $\Phi_{as}$
multiplied by polynomials that are orthogonal over the weight function (\ref{as-dis}). For LCDAs in Table \ref{tabDA-def}, we give their conformal expansions:
\begin{eqnarray}
\Phi_3(x_i)&=&120x_1x_2x_3[\phi_3^0+\phi_3^-(x_1-x_2)+\phi_3^+(1-3x_3)+...],\nonumber\\
 T_1(x_i)&=&120x_1x_2x_3[t_1^0+t_1^-(x_1-x_2)+t_1^+(1-3x_3)+...]\label{contwist3}
\end{eqnarray}
for twist-$3$ and
\begin{eqnarray}
\Phi_4(x_i)&=&24x_1x_2[\phi_4^0+\phi_4^-(x_1-x_2)+\phi_4^+(1-5x_3)+...],\nonumber\\
\Psi_4(x_i)&=&24x_1x_3[\psi_4^0+\psi_4^-(x_1-x_3)+\psi_4^+(1-5x_2)+...],\nonumber\\
\Xi_4(x_i)&=&24x_2x_3[\xi_4^0+\xi_4^-(x_2-x_3)+\xi_4^+(1-5x_1)+...],\nonumber\\
{\Xi'}_4(x_i)&=&24x_2x_3[{\xi'}_4^0+{\xi'}_4^-(x_2-x_3)+{\xi'}_4^+(1-5x_1)+...],\nonumber\\
T_2(x_i)&=&24x_1x_2[t_2^0+t_2^-(x_1-x_2)+t_2^+(1-5x_3)+...]\label{contwist4}
\end{eqnarray}
for twist-$4$ and
\begin{eqnarray}
\Phi_5(x_i)&=&6x_3[\phi_5^0+\phi_5^-(x_1-x_2)+\phi_5^+(1-2x_3)+...],\nonumber\\
\Psi_5(x_i)&=&6x_2[\psi_5^0+\psi_5^-(x_1-x_3)+\psi_5^+(1-2x_2)+...],\nonumber\\
\Xi_5(x_i)&=&6x_1[\xi_5^0+\xi_5^-(x_2-x_3)+\xi_5^+(1-2x_1)+...],\nonumber\\
{\Xi'}_5(x_i)&=&6x_1[{\xi'}_5^0+{\xi'}_5^-(x_2-x_3)+{\xi'}_5^+(1-2x_1)+...],\nonumber\\
T_5(x_i)&=&6x_3[t_5^0+t_5^-(x_1-x_2)+t_5^+(1-2x_3)+...]\label{contwist5}
\end{eqnarray}
for twist-$5$, and
\begin{eqnarray}
\Phi_6(x_i)&=&2[\phi_6^0+\phi_6^-(x_1-x_2)+\phi_6^+(1-3x_3)+...],\nonumber\\
 T_6(x_i)&=&2[t_6^0+t_6^-(x_1-x_2)+t_6^+(1-3x_3)+...] \label{contwist6}
\end{eqnarray}
for twist-$6$. Up to now there are altogether $42$ parameters which need to be determined.

To the next-to-leading order, the normalization of the $\Sigma$ baryon LCDAs is determined by the matrix element of the nonlocal three-quark operator expanded at the zero point. The decomposition of the matrix element is
\begin{eqnarray}
\langle0|\epsilon^{ijk}u^i_\alpha(a_1z)u^j_\beta(a_2z)s^k_\gamma(a_3z)|\Sigma(P)\rangle\nonumber
=\langle0|\epsilon^{ijk}u^i_\alpha(a_1z)u^j_\beta(a_2z)s^k_\gamma(a_3z)|\Sigma(P)\rangle\nonumber\\
+z_\lambda\langle0|[\epsilon^{ijk}u^i_\alpha(a_1z)\stackrel{\leftrightarrow}{D}u^j_\beta(a_2z)]s^k_\gamma(a_3z)|\Sigma(P)\rangle\nonumber\\
+z_\lambda\langle0|\epsilon^{ijk}u^i_\alpha(a_1z)u^j_\beta(a_2z)[\vec Ds^k_\gamma(a_3z)]|\Sigma(P)\rangle.
\end{eqnarray}

The Lorentz decomposition of the matrix element can be expressed explicitly as
\begin{eqnarray}
4\langle0|\epsilon^{ijk}s^i_\alpha(0)s^j_\beta(0)q^k_\gamma(0)|\Sigma(P)\rangle=\mathcal{V}^0_1(\!\not\!
PC)_{\alpha\beta}(\gamma_5\Sigma)_\gamma+\mathcal{V}^0_3(\gamma_\mu
C)_{\alpha\beta}(\gamma_\mu\gamma_5\Sigma)_\gamma\nonumber\\
+\mathcal{T}^0_1(P^\nu i\sigma_{\mu\nu}C)_{\alpha \beta}(\gamma^\mu\gamma_5\Sigma)_\gamma+\mathcal{T}^0_3M(\sigma_{\mu\nu}C)_{\alpha
\beta}(\sigma^{\mu\nu}\gamma_5\Sigma)_\gamma
\end{eqnarray}
for the matrix element of the leading order, and
\begin{eqnarray}
&&4\langle0|\epsilon^{ijk}u^i_\alpha(a_1z)u^j_\beta(a_2z)[\vec Ds^k_\gamma(a_3z)]|\Sigma(P)\rangle\nonumber\\
&&=\mathcal{V}_1^s(\!\not\!
PC)_{\alpha\beta}(\gamma_5\Sigma)_\gamma+\mathcal{V}_2^0M(\!\not\!
PC)_{\alpha\beta}(\gamma_\lambda\gamma_5\Sigma)_\gamma+\mathcal{V}_3^sP_\lambda M(\gamma_\mu
C)_{\alpha\beta}(\gamma_\mu\gamma_5\Sigma)_\gamma\nonumber\\
&&+\mathcal{V}_4^0M^2(\gamma_\lambda C)_{\alpha\beta}(\gamma_5\Sigma)_\gamma
+\mathcal{V}_5^0M^2(\gamma_\mu C)_{\alpha\beta}(i\sigma_{\mu\lambda}\gamma_5\Sigma)_\gamma
+\mathcal{T}_1^sP_\lambda(P^\nu i\sigma_{\mu\nu}C)_{\alpha \beta}(\gamma^\mu\gamma_5\Sigma)_\gamma\nonumber\\
&&+\mathcal{T}_2^0M(P^\nu i\sigma_{\lambda\nu}C){\gamma_5\Sigma}_\gamma+\mathcal{T}_3^sMP_\lambda(\sigma_{\mu\nu}C)_{\alpha
\beta}(\sigma^{\mu\nu}\gamma_5\Sigma)_\gamma+\mathcal{T}_4^0M(P_\nu\sigma_{\mu\nu}C)_{\alpha
\beta}(\sigma^{\mu\lambda}\gamma_5\Sigma)_\gamma\nonumber\\
&&+\mathcal{T}_5^0M^2(i\sigma_{\mu\lambda}C)_{\alpha
\beta}(\gamma^\mu\gamma_5\Sigma)_\gamma
+\mathcal{T}_7^0M^2(\sigma_{\mu\nu}C)_{\alpha
\beta}(\sigma^{\mu\nu}\gamma_\lambda\gamma_5\Sigma)_\gamma,
\end{eqnarray}
\begin{eqnarray}
&&4\langle0|[\epsilon^{ijk}u^i_\alpha(a_1z)\stackrel{\leftrightarrow}{D}u^j_\beta(a_2z)]s^k_\gamma(a_3z)|\Sigma(P)\rangle\nonumber\\
&&=\mathcal{S}_1^uP_\lambda MC_{\alpha\beta}(\gamma_5\Sigma)_\gamma
+\mathcal{S}_2^0M^2C_{\alpha\beta}(\gamma_\lambda\gamma_5\Sigma)_\gamma+\mathcal{P}_1^uP_\lambda M(\gamma_5
C)_{\alpha\beta}\Sigma_\gamma+\mathcal{P}_2^0M^2(\gamma_5 C)_{\alpha\beta}(\gamma_\lambda\Sigma)_\gamma\nonumber\\
&&+\mathcal{A}_1^uP_\lambda(\!\not\!P\gamma_5 C)_{\alpha\beta}\Sigma_\gamma+\mathcal{A}_2^0M(\!\not\!P\gamma_5 C)_{\alpha\beta}{\gamma_\lambda\Sigma}_\gamma+\mathcal{A}_3^0P_\lambda M(\gamma_\mu\gamma_5 C)_{\alpha\beta}{\gamma_\mu\Sigma}_\gamma\nonumber\\
&&+\mathcal{A}_4^0M^2(\gamma_\lambda\gamma_5 C)_{\alpha\beta}\Sigma_\gamma+\mathcal{A}_5^0 M^2(\gamma_\mu\gamma_5 C)_{\alpha\beta}{i\sigma_{\mu\lambda}\Sigma}_\gamma
\end{eqnarray}
for the next leading order expansion. There are altogether $24$ nonperturbative parameters in the expressions. However, we need not so many free parameters because there are some constraints to reduce the freedom of the coefficients. It is noticed that all the parameters defined above are not independent and can be reduced with the help of the motion of equation, which can be seen in Appendix \ref{app:1}.

Choosing $\mathcal{V}_1^0,\mathcal{V}_3^0,\mathcal{V}_1^s,\mathcal{V}_3^s,\mathcal{T}_1^0,\mathcal{T}_3^0,\mathcal{T}_1^s,\mathcal{T}_3^s,\mathcal{A}_1^u,
\mathcal{A}_3^u,\mathcal{S}_1^u,\mathcal{P}_0^2$ as the independent parameters, the other ones can be expressed with them:
\begin{eqnarray}
&\mathcal{V}_2^0=\frac14(\mathcal{V}_1^s-2\mathcal{V}_3^s),&\mathcal{V}_4^0=\frac{1}{16}(4\mathcal{V}_1^0-4\mathcal{V}_3^0-3\mathcal{V}_1^s+2\mathcal{V}_3^s),\nonumber\\
&\mathcal{V}_5^0=\frac{1}{48}(-4\mathcal{V}_1^0+4\mathcal{V}_3^0+3\mathcal{V}_1^s-50\mathcal{V}_3^s), &\mathcal{T}_2^0=\frac{1}{10}(3\mathcal{S}_1^u-3\mathcal{T}_1^0+6\mathcal{T}_3^0+2\mathcal{T}_1^s-2\mathcal{T}_3^s),\nonumber\\
&\mathcal{T}_4^0=\frac{1}{10}(\mathcal{S}_1^u-\mathcal{T}_1^s+2\mathcal{T}_3^0+4\mathcal{T}_1^s-14\mathcal{T}_3^s),&\mathcal{T}_5^0=-\mathcal{T}_3^s,\nonumber\\
&\mathcal{T}_7^0=\frac{1}{30}(5\mathcal{P}_2^0-\mathcal{S}_1^u+\mathcal{T}_1^0-12\mathcal{T}_3^0-4\mathcal{T}_1^s+24\mathcal{T}_3^s),&\mathcal{A}_2^0
=\frac{1}{4}(4\mathcal{A}_3^u-4\mathcal{V}_3^0-\mathcal{V}_1^s+6\mathcal{V}_3^s),\nonumber\\
&\mathcal{A}_4^0=\frac{1}{16}(-4\mathcal{A}_1^u-8\mathcal{A}_3^u+4\mathcal{V}_3^0+\mathcal{V}_1^s-6\mathcal{V}_3^s),&\mathcal{A}_5^0=\frac{1}{48}(4\mathcal{V}_1^0
+20\mathcal{V}_3^0+3\mathcal{V}_1^s+14\mathcal{V}_3^s),\nonumber\\
&\mathcal{S}_2^0=\frac{1}{10}(-10\mathcal{P}_2^0+3S_1^u+7\mathcal{T}_1^0+6\mathcal{T}_3^0+2\mathcal{T}_1^s-12\mathcal{T}_3^s),&\nonumber\\
&\mathcal{P}_1^u=\frac{1}{5}(-\mathcal{S}_1^u+\mathcal{T}_1^0-12\mathcal{T}_3^0-4\mathcal{T}_1^s+24\mathcal{T}_3^s).&
\end{eqnarray}

Recall the relations of the leading order, there are altogether $12$ parameters to be determined. To this end, we introduce the additional eight decay constants defined by the following matrix elements of a three-quark operator with a covariant derivative:
\begin{eqnarray}
&& \langle{0}| \epsilon^{ijk} \left[u^i(0) C \!\not\!{z} u^j(0)\right] \, \gamma_5 \!\not\!{z} (iz\vec{D}s^k)(0)| {\Sigma(P)}\rangle = f_{\Sigma}
V_1^s(P\cdot z)^2 \!\not\!{z} \Sigma(P)_\gamma\,, \nonumber \\
&& \langle{0}| \epsilon^{ijk} \left[u^i(0) C \!\not\!{z}\gamma_5 iz\stackrel{\leftrightarrow}{D}u^j(0)\right] \!\not\!{z} s^k(0)| {\Sigma(P)}\rangle =-f_{\Sigma}
A_1^u(P\cdot z)^2 \!\not\!{z} \Sigma(P)_\gamma\,, \nonumber \\
&& \langle{0}| \epsilon^{ijk} \left[u^i(0) C \gamma^u u^j(0)\right] \, \gamma_5 \!\not\!{z}\gamma^u (iz\vec{D}s^k)(0)| {\Sigma}\rangle = \lambda_1f_1^s
(P\cdot z)M \!\not\!{z} \Sigma(P)_\gamma\,, \nonumber \\
&& \langle{0}| \epsilon^{ijk} \left[u^i(0) C \sigma_{\mu\nu} u^j(0)\right] \, \gamma_5 \!\not\!{z}\sigma_{\mu\nu} (iz\vec{D}s^k)(0)| {\Sigma}\rangle =-\lambda_2
f_2^s(P\cdot z)M \!\not\!{z} \Sigma(P)_\gamma\,, \nonumber \\
&& \langle{0}| \epsilon^{ijk} \left[u^i(0) C \gamma_\mu\gamma_5 iz\stackrel{\leftrightarrow}{D}u^j(0)\right] \!\not\!{z}\gamma^\mu s^k(0)| {\Sigma(P)}\rangle =-\lambda_1
f_1^u(P\cdot z)M\!\not\!{z} \Sigma(P)_\gamma\,, \nonumber \\
&& \langle{0}| \epsilon^{ijk} \left[u^i(0)iP^\nu C\sigma^{\mu\nu} u^j(0)\right] \, \gamma_5 \!\not\!{z} (iz\vec{D}s^k)(0)| {\Sigma(P)}\rangle =-\lambda_3
f_3^s(P\cdot z)M^2 \!\not\!{z} \Sigma(P)_\gamma\,, \nonumber \\
&& \langle{0}| \epsilon^{ijk} \left[u^i(0) C iz\stackrel{\leftrightarrow}{D}u^j(0)\right] \gamma_5 s^k(0)| {\Sigma(P)}\rangle =\mathcal{S}_1^u(P\cdot z)M\Sigma(P)-\mathcal{S}_2^0M^2(\!\not\!{z}\Sigma(P))_\gamma\,,\nonumber\\
&& \langle{0}| \epsilon^{ijk} \left[u^i(0) C iz\stackrel{\leftrightarrow}{D} \gamma_5 u^j(0)\right] s^k(0)| {\Sigma(P)}\rangle =\mathcal{P}_1^u(P\cdot z)M\Sigma(P)+\mathcal{P}_2^0M^2(\!\not\!{z}\Sigma(P))_\gamma\,. \label{def-nonlocal2}
\end{eqnarray}
It is noticed that each of the last two matrix element have two different Lorentz structures which permit us to get two different sum rules; whereas the calculations also indicate that the sum rules from the last two ones are the same, so we can get the necessary equations from the two different sum rules.

We also need another four decay constant defined by the leading order local operator matrix element which has been calculated in the previous paper \cite{DAs}
\begin{eqnarray}
&& \langle{0}| \epsilon^{ijk} \left[u^i(0) C \!\not\!{z} u^j(0)\right] \, \gamma_5 \!\not\!{z} s^k(0)| {P}\rangle = f_{\rm \Sigma}
P\cdot z \!\not\!{z} \Sigma(P)\,, \nonumber \\
&&\langle{0}| \epsilon^{ijk} \left[u^i(0) C\gamma_\mu u^j(0)\right]\, \gamma_5 \gamma^\mu s^k(0)| {P}\rangle = \lambda_1 M \Sigma(P) \,,
\nonumber \\
&&\langle{0}| \epsilon^{ijk} \left[u^i(0) C\sigma_{\mu\nu} u^j(0)\right] \, \gamma_5 \sigma^{\mu\nu} s^k(0)|{P}\rangle = \lambda_2 M \Sigma(P) \,,
\nonumber\\
&&\langle{0}| \epsilon^{ijk} \left[u^i(0) Ciq^\nu\sigma_{\mu\nu}u^j(0)\right]\,\gamma_5\gamma_\mu s^k(0)|{P}\rangle=\lambda_3 M\!\not\!{q}\Sigma(P)
\,.\label{def-nonlocal}
\end{eqnarray}
A simple calculation gives the relations between the local nonperturbative parameters $\mathcal V_i^{0,s}, \mathcal A_i^{u}$, $\mathcal T_i^{0,s}$, $\mathcal P_2^{0}$, $\mathcal S_1^{u}$ and the decay constants defined in Eqs. (\ref{def-nonlocal2}) and (\ref{def-nonlocal}):
\begin{eqnarray}
f_{\Sigma}&=&\mathcal V_1^0,\hspace{3.8cm}\lambda_1=\mathcal V_1^0-4\mathcal V_3^0,\nonumber\\
\lambda_2&=&6\mathcal T_1^0-24\mathcal T_3^0,\hspace{2.1cm}\lambda_3=3\mathcal T_1^0-6\mathcal T_3^0,\nonumber\\
f_\Sigma V_1^s&=&\mathcal V_1^s,\hspace{3.8cm}f_\Sigma A_1^u=\mathcal A_1^u,\nonumber\\
\lambda_1 f_1^s&=&-\mathcal V_1^s+4\mathcal V_3^s+2\mathcal V_2^0,\hspace{1.0cm}\lambda_2f_2^s=6\mathcal T_1^s-2\mathcal T_2^0-24 \mathcal T_3^s-8\mathcal T_4^0,\nonumber\\
\lambda_1 f_1^u&=&\mathcal A_1^u+2\mathcal A_2^0+4\mathcal A_3^u,\hspace{1.1cm}\lambda_3f_3^s=3\mathcal T_3^s+\mathcal T_2^0-6\mathcal T_3^s-\mathcal T_4^0+4\mathcal T_5^0+12\mathcal T_7^0.
\end{eqnarray}
We also noticed that $\mathcal S_1^u$ and $\mathcal P_2^0$ are defined directly by the matrix element and can be determined by the following method. Up to now we can express all the independent parameters by the nonperturbative decay constants defined in Eqs. (\ref{def-nonlocal2}) and (\ref{def-nonlocal}):
\begin{eqnarray}
\mathcal V_1^0&=&f_\Sigma, \hspace{1cm}\mathcal V_3^0=\frac{1}{4}(f_\Sigma-\lambda_1),\hspace{1cm}\mathcal V_1^s=f_\Sigma V_1^s,\hspace{1cm}
\mathcal V_3^s=\frac{1}{2}f_1^s\lambda_1,\nonumber\\
\mathcal T_1^0&=&\frac{1}{6}(-\lambda_2+4\lambda_3),\nonumber\\
\mathcal T_3^0&=&\frac{1}{12}(-\lambda_2+2\lambda_3),\nonumber\\
\mathcal T_1^s&=&\frac{62}{33}P_2^0-\frac{13}{22}S_1^u+\frac{31}{99}\lambda_2-\frac{19}{66}f_2^s\lambda_2-\frac{85}{198}\lambda_3-\frac{31}{33}f_3^s\lambda_3,\nonumber\\
\mathcal T_3^s&=&\frac{4}{11}P_2^0-\frac{5}{22}S_1^u+\frac{2}{33}\lambda_2-\frac{3}{22}f_2^s\lambda_2-\frac{1}{22}\lambda_3-\frac{2}{11}f_3^s\lambda_3,\nonumber\\
\mathcal A_1^u&=&f_\Sigma A_1^u,\hspace{1cm}A_3^u=\frac{1}{12}(f_\Sigma-2f_1A_1^u+f_\Sigma V_1^s-\lambda_1-3f_1^s\lambda_1+2f_1^u\lambda_1).
\end{eqnarray}

Further calculation shows that coefficients in Eqs. (\ref{contwist3})-(\ref{contwist6}) can be expressed to the next-to-leading order conformal spin accuracy as
\begin{eqnarray}
\phi_3^0&=&\phi_6^0={\mathcal V}_1^0,\hspace{4cm}\psi_4^0=\psi_5^0=2{\mathcal V}_3^0,\nonumber\\
\phi_4^0&=&\phi_5^0={\mathcal V}_1^0-2{\mathcal V}_3^0,\hspace{2.8cm}t_1^0={\xi'}_4^0=-\xi_5^0={\mathcal T}_1^0,\nonumber\\
t_2^0&=&t_5^0=\xi_4^0=-{\xi'}_5^0={\mathcal T}_1^0-4{\mathcal T}_3^0,\hspace{0.6cm}-\xi_5^0=t_6^0={\mathcal T}_1^0
\end{eqnarray}
for leading order and
\begin{eqnarray}
\phi_3^+&=&\frac{21}{6}{\mathcal V}_1^0-{\mathcal V}_1^s,\hspace{4.6cm} \phi_6^+=2{\mathcal V}_1^0-6{\mathcal V}_1^s+12{\mathcal V}_2^0-12{\mathcal V}_4^0-12{\mathcal V}_5^0,\nonumber\\
t_1^+&=&\frac{1}{2}({7\mathcal T}_1^0-21{\mathcal T}_1^s),\hspace{4.0cm} \phi_4^+=\frac{3}{2}({\mathcal V}_1^0-2{\mathcal V}_3^0)-\frac{15}{2}({\mathcal V}_1^s-2{\mathcal V}_2^0-2{\mathcal V}_3^s),\nonumber\\
t_6^+&=&2{\mathcal T}_1^0-6{\mathcal T}_1^s+12{\mathcal T}_4^0-12{\mathcal T}_5^0,\hspace{1.8cm}t_1^-=t_2^-=t_5^-=t_6^-=0\nonumber\\
\phi_5^+&=&5{\mathcal V}_1^0-10{\mathcal V}_3^0-10{\mathcal V}_1^s-20{\mathcal V}_4^0+20{\mathcal V}_5^0+20{\mathcal V}_3^s,\nonumber\\
t_5^+&=&5{\mathcal T}_1^0+20{\mathcal T}_3^0-10{\mathcal T}_1^s-20{\mathcal T}_5^0-80{\mathcal T}_7^0,\nonumber\\
t_2^+&=&\frac{3}{2}({\mathcal T}_1^0-4{\mathcal T}_3^0)-15({\mathcal T}_1^s-2{\mathcal T}_4^0-4{\mathcal T}_3^s),\nonumber\\
\phi_3^-&=&-\frac{21}{2}{\mathcal A}_1^u,\hspace{5.2cm}
\phi_6^-=-6({\mathcal A}_1^u+2{\mathcal A}_2^0+2{\mathcal A}_4^0+2{\mathcal A}_5^0),\nonumber\\
\phi_4^-&=&\frac{15}{2}({\mathcal A}_1^u+2{\mathcal A}_2^0-2{\mathcal A}_3^u),\hspace{2.8cm}
\phi_5^-=-10({\mathcal A}_1^u+2{\mathcal A}_4^0-2{\mathcal A}_5^0+2{\mathcal A}_3^u),\nonumber\\
\psi_4^+&=&\frac{15}{2}({\mathcal V}_3^s-{\mathcal A}_3^u)-\frac{9}{2}{\mathcal V}_3^0,\hspace{3cm}
\psi_4^-=\frac{15}{2}({\mathcal V}_3^0-{\mathcal A}_3^u)-\frac{45}{2}{\mathcal V}_3^s,\nonumber\\
\psi_5^+&=&40({\mathcal V}_3^s+2{\mathcal V}_5^0-{\mathcal A}_3^u+2{\mathcal A}_5^0),\hspace{2cm}
\psi_5^-=10({\mathcal V}_3^0-3{\mathcal V}_3^s-6{\mathcal V}_5^0-{\mathcal A}_3^u+2{\mathcal A}_5^0),\nonumber\\
{\xi'}_4^+&=&3(2{\mathcal S}_1^u-2{\mathcal P}_1^u-{\mathcal T}_1^s+2{\mathcal T}_2^0),\hspace{2cm}
{\xi'}_4^-=-3({\mathcal S}_1^u-{\mathcal P}_1^u)-3{\mathcal T}_1^0+9({\mathcal T}_1^s-{\mathcal T}_2^0),\nonumber\\
\xi_4^+&=&6({\mathcal S}_1^u+{\mathcal P}_1^u)+\frac{3}{10}({\mathcal T}_1^0-4{\mathcal T}_3^0+10{\mathcal T}_2^0-5{\mathcal T}_1^s+8{\mathcal T}_3^s),\nonumber\\
\xi_4^-&=&-3({\mathcal S}_1^u+{\mathcal P}_1^u)-\frac{9}{10}(\frac{13}{10}{\mathcal T}_1^0-\frac{26}{5}{\mathcal T}_3^0+10{\mathcal T}_2^0
-5{\mathcal T}_1^s+8{\mathcal T}_3^s),\nonumber\\
\xi_5^+&=&20({\mathcal T}_3^0-2{\mathcal T}_7^0)-15({\mathcal T}_1^s+2{\mathcal T}_5^0+{\mathcal T}_2^0-{\mathcal T}_4^0)+5({\mathcal S}_1^u
+{\mathcal P}_1^u-2{\mathcal S}_2^0+2{\mathcal P}_2^0),\nonumber\\
\xi_5^-&=&-5{\mathcal T}_1^0-6({\mathcal T}_3^0-2{\mathcal T}_7^0)+45({\mathcal T}_1^s+2{\mathcal T}_5^0+{\mathcal T}_2^0-{\mathcal T}_4^0)
+5({\mathcal S}_1^u+{\mathcal P}_1^u-2{\mathcal S}_2^0+2{\mathcal P}_2^0),\nonumber\\
{\xi'}_5^+&=&40({\mathcal T}_3^0-2{\mathcal T}_7^0)-30({\mathcal T}_1^s+2{\mathcal T}_5^0+{\mathcal T}_2^0-{\mathcal T}_4^0)+5({\mathcal S}_1^u
-{\mathcal P}_1^u-2{\mathcal S}_2^0-2{\mathcal P}_2^0),\nonumber\\
{\xi'}_5^-&=&-5({\mathcal T}_1^0-4{\mathcal T}_3^0)-120({\mathcal T}_3^0-2{\mathcal T}_7^0)+90({\mathcal T}_1^s+2{\mathcal T}_5^0+{\mathcal T}_2^0
-{\mathcal T}_4^0)\nonumber\\
&&+5({\mathcal S}_1^u-{\mathcal P}_1^u-2{\mathcal S}_2^0-2{\mathcal P}_2^0)
\end{eqnarray}
for the next-to-leading order.

\section{numerical analysis of the sum rules for the nonperturbative parameters}\label{sec:sumrule}

The nonperturbative parameters appearinf in the above section can be determined with QCD sum rules \cite{SVZ}. The QCD sum rule approach is a well-used tool to estimate unknown physical parameters which are connected with the nonperturbative effects in low energy scale of strong interaction. Early in the 1980s the QCD sum rules were used to calculate the moments of the meson and baryon LCDAs \cite{Cheryak-sum}. The detailed analysis of the sum rules for $V_1^s$ is presented in Appendix \ref{app:2} as an example for the approach. Analysis of other sum rules are the same as the example. In this section we only present the explicit expressions of the parameters from this method. It is noticed that the parameters related with the leading order conformal spin expansion have been obtained in Ref. \cite{DAs}. Herein we only present the next-to-leading order ones. The sum rules are as follows:
\begin{itemize}
\item
The sum rule for $V_1^s$ is
\begin{eqnarray}
2f_\Sigma^2V_1^se^{-M^2/M_B^2}=\int_{m_s^2}^{s_0}e^{-s/M_B^2}\rho(s)ds+\Pi^{cond.},\label{sumruleofv1s}
\end{eqnarray}
where
\begin{eqnarray}
\rho(s)&=&\frac{1}{5\times3\times2^5\pi^4}s(1-x)^5(1+2x)+\frac{\langle g^2G^2\rangle}{3\times2^6\pi^4}\frac{1}{s}x^3(1-x)\nonumber\\
&&+\frac{\langle g^2G^2\rangle}{3^2\times2^6\pi^4}\frac{1}{s}x(1-x)^2(1-4x),
\end{eqnarray}
and
\begin{equation}
\Pi^{cond.}=\frac{m(m_0^2-2m_s^2)}{3^2\times2^3\pi^2}\langle \bar ss\rangle\frac{1}{M_B^2}-\frac{m_s}{3^2\times2^4\pi^2}\langle \bar sg\cdot \sigma Gs\rangle\frac{1}{M_B^2}(1+\frac{m_s^2}{M_B^2}),
\end{equation}
where $x=m_s^2/s$, $m_s$ is the strange quark mass, $M$ is the mass of $\Sigma$ and $M_B^2$ is the Borel parameter.
\item
The sum rule for $A_1^u$ is
\begin{eqnarray}
2f_\Sigma^2A_1^ue^{-M^2/M_B^2}=\int_{m_s^2}^{s_0}e^{-s/M_B^2}\rho(s)ds+\Pi^{cond.},
\end{eqnarray}
where
\begin{equation}
\rho(s)=\frac{\langle g^2G^2\rangle}{3^2\times2^3\pi^4}\frac{1}{s}x(1-x)^3,
\end{equation}
and
\begin{equation}
\Pi^{cond.}=\frac{\langle \bar s\sigma\cdot Gs\rangle}{3\times2^4\pi^2}\frac{m_s}{M_B^2}-\frac{\langle \bar s\sigma\cdot Gs\rangle}{3^2\times2^3\pi^2}\frac{m_s^3}{M_B^4}.
\end{equation}
\item
The sum rule for $f_1^s$ is
\begin{eqnarray}
\lambda_1^2M^2f_1^{s}e^{-M^2/M_B^2}=\int_{m_s^2}^{s_0}e^{-s/M_B^2}\rho^{}(s)ds+\Pi^{cond.},
\end{eqnarray}
where
\begin{eqnarray}
\rho^{}(s)&=&-\frac{s^2}{5\times3\times2^6\pi^4}\{\frac{1}{2}(1-x)(9-21x+119x^2-61x^3+14x^4)\nonumber\\
&&+30x^2\ln x\}+\frac{\langle g^2G^2\rangle}{3^2\times2^8\pi^4}(1-x)(1-25x+32x^2),
\end{eqnarray}
\begin{eqnarray}
\Pi^{cond.}&=&\frac{m_s M_B^2}{24\pi^2}\langle\bar ss\rangle+\frac{m_s(m_0^2-2m_s^2)}{48\pi^2}\langle\bar ss\rangle-\frac{2}{3}\langle\bar qq\rangle^2e^{-m_s^2/M_B^2}(1-\frac{m_0^2}{M_B^2}\nonumber\\
&&-2\frac{m_0^2m_s^2}{M_B^4})+\frac{m_s\langle\bar sg\sigma\cdot Gs\rangle}{3^2\times2^5\pi^2}(3-\frac{m_s^2}{M_B^2}).
\end{eqnarray}
\item
The independent sum rule for $f_2^s$ is
\begin{eqnarray}
-\lambda_2^2M^2f_2^{s}e^{-M^2/M_B^2}=\int_{m_s^2}^{s_0}e^{-s/M_B^2}\rho^{}(s)ds+\Pi^{cond.},
\end{eqnarray}
where
\begin{eqnarray}
\rho^{}(s)&=&-\frac{s^2}{5\times3\times2^4\pi^4}\{[(1-x)(16-79x+31x^2-39x^3+11x^4)\nonumber\\
&&-60x^2\ln x]\}-\frac{\langle g^2G^2\rangle}{3^2\times2^5\pi^4}(1-x)(19+223x-233x^2),
\end{eqnarray}
\begin{eqnarray}
\Pi^{cond.}&=&-\frac{m_sl}{3\pi^2}\langle\bar ss\rangle( M_B^2-\frac{1}{6}(m_0^2-2m_s^2))+\frac{m_s\langle\bar sg\sigma\cdot Gs\rangle}{12\pi^2}(2+\frac{m_s^2}{M_B^2}).
\end{eqnarray}
\item
The independent sum rule for $f_1^u$ is
\begin{eqnarray}
-\lambda_1^2M^2f_1^{u}e^{-M^2/M_B^2}=\int_{m_s^2}^{s_0}e^{-s/M_B^2}\rho^{}(s)ds+\Pi^{cond.},
\end{eqnarray}
where
\begin{eqnarray}
\rho^{}(s)&=&\frac{s^2}{5\times3\times 2^7\pi^4}\{[(1-x)(3-27x-47x^2+13x^3-2x^4)\nonumber\\
&&-60x^2\ln x]\}+\frac{\langle g^2G^2\rangle}{3^2\times2^8\pi^4}(1-x)^2(5-4x),
\end{eqnarray}
\begin{eqnarray}
\Pi^{cond.}&=&\frac{m_s}{3\times2^4\pi^2}\langle\bar ss\rangle(2M_B^2-m_0^2+2m_s^2)-\frac{5m_s\langle\bar sg\sigma\cdot Gs\rangle}{3^2\times2^5\pi^2}(3+2\frac{m_s^2}{M_B^2}).
\end{eqnarray}
\item
The independent sum rule for $f_3^s$ is
\begin{eqnarray}
-\lambda_3^2M^3f_3^{s}e^{-M^2/M_B^2}=\int_{m_s^2}^{s_0}e^{-s/M_B^2}\rho^{}(s)ds+\Pi^{cond.},
\end{eqnarray}
where
\begin{eqnarray}
\rho^{}(s)&=&-\frac{m_s}{5\times2^8\pi^4}s^2\{(1-x)(9-51x-11x^2-11x^3+4x^4)-60x^2\ln(x)]\}\nonumber\\
&&+\frac{m_s\langle g^2G^2\rangle}{5\times2^9\pi^4}\frac{1}{s}(1-x)(13-29x+29x^2-3x^3)\nonumber\\
&&-\frac{m_s\langle g^2G^2\rangle}{3\times2^9\pi^4}\{(1-x)(131-79x+20x^2)+72\ln x\},
\end{eqnarray}
\begin{eqnarray}
\Pi^{cond.}&=&\frac{(m_0^2-2m_s^2)M_B^2}{2^5\pi^2}\langle\bar ss\rangle-m_s\langle\bar qq\rangle^2e^{-m_s^2/M_B^2}(1+\frac{m_0^2}{M_B^2}-\frac{m_0^2m_s^2}{M_B^4})\nonumber\\
&&+\frac{M_B^2}{3\times2^5\pi^2}\langle \bar sg\sigma\cdot Gs\rangle-\frac{52m_s^2}{3^2\times2^7\pi^2}\langle \bar sg\sigma\cdot Gs\rangle.
\end{eqnarray}
\item
The sum rules of $\mathcal S_1^u$ and $\mathcal S_2^0$ are
\begin{eqnarray}
f^*M\mathcal S_1^ue^{-M^2/M_B^2}=\int_{m_s^2}^{s_0}e^{-s/M_B^2}\rho^{(1)}(s)ds+\Pi^{(1)cond.},
\end{eqnarray}
where
\begin{eqnarray}
\rho^{(1)}(s)&=&\frac{1}{5\times2^8\pi^4}s^2[-(1-x)(3-27x-47x^2+13x^3-2x^4)+60x\ln x]\nonumber\\
&&-\frac{\langle g^2G^2\rangle}{3\times2^9\pi^4}(1-x)^2(1+2x)+\frac{\langle g^2G^2\rangle}{3\times2^8\pi^4}(1-x)^3,
\end{eqnarray}
\begin{eqnarray}
\Pi^{(1)cond.}&=&-\frac{m_s}{2^5\pi^2}\langle\bar ss\rangle(4M_B^2-(m_0^2-2m_s^2))-\frac{m_s\langle\bar sg\sigma\cdot Gs\rangle}{3^2\times2^6\pi^2}(3-2\frac{m_s^2}{M_B^2}),
\end{eqnarray}
and
\begin{eqnarray}
f^*M^2(\mathcal S_1^u-2\mathcal S_2^0)e^{-M^2/M_B^2}=\int_{m_s^2}^{s_0}e^{-s/M_B^2}\rho^{(2)}(s)ds+\Pi^{(2)cond.},
\end{eqnarray}
with
\begin{eqnarray}
\rho^{(2)}(s)&=&\frac{1}{2^8\pi^4}s^2[(1-x)(3+47x+11x^2-x^3)+12x(2+3x)\ln x]\nonumber\\
&&+\frac{m_s\langle g^2G^2\rangle}{2^{9}\pi^4}[(1-x)^2-\frac{(1-x)(2+5x-x^2)+6x\ln x}{x}]\nonumber\\
&&-\frac{m_s\langle g^2G^2\rangle}{3\times2^9\pi^4}[2\frac{(1-x)^3}{x}+(1-x)(3-x)+2\ln x],
\end{eqnarray}
\begin{eqnarray}
\Pi^{(2)cond.}&=&\frac{M_B^4}{8\pi^2}\langle\bar ss\rangle+\frac{3(m_0^2-2m_s^2)}{32\pi^2}M_B^2\langle\bar ss\rangle+\frac{\langle\bar sg\sigma\cdot Gs\rangle}{3\times2^5\pi^2}(m_s^2+M_B^2).
\end{eqnarray}
\end{itemize}

The calculation also shows that sum rules for $P_1^u$ and $P_2^0$ are the same as that for $S_1^u$ and $ S_2^0$. Therefore we do not show them explicitly in the text.

In addition, we need to calculate the parameter $f^*$ to get the numerical results of $S_1^u$ and $S_2^0$. The parameter $f^*$ is defined by the following matrix element:
\begin{equation}
\langle{0}| \epsilon^{ijk} \left[u^i(0) C u^j(0)\right] \, \gamma_5 \!\not\!{z} s^k(0)| \Sigma(P)\rangle = f^*\not\!{z}\Sigma.
\end{equation}
In compliance with the standard procedure of QCD sum rules, we arrive at the final result:
\begin{eqnarray}
2f^{*2}e^{-M^2/M_B^2}=\int_{m_s^2}^{s_0}e^{-s/M_B^2}\rho(s)ds+\Pi^{cond.},
\end{eqnarray}
with
\begin{eqnarray}
\rho(s)&=-&\frac{3}{2^5\pi^4}[(1-x)(3-x)+2\ln x]-\frac{\langle g^2G^2\rangle}{3\times2^{9}\pi^4}(1-x)(-3+5x),\\
\Pi^{cond.}&=&\frac{2}{3}\langle \bar qq\rangle^2e^{-\frac{m_s^2}{M_B^2}}+\frac{m_s\langle\bar ss\rangle}{3\times2^3\pi^2}(3M_B^2-m_0^2+2m_s^2).
\end{eqnarray}

In fact, there are two different Lorentz structures which may give independent sum rules for most of the above parameters in the calculation. In practice we choose the proper ones which may contain more information of the hadrons and have good Borel working windows. Furthermore, in order to cancel uncertainties from auxiliary parameters such as Borel mass and the threshold $s_0$ as far as possible, we use the sum rules other than the central values in numerical analysis. For example, when analyzing Eq. (\ref{sumruleofv1s}), the parameter $f_\Sigma^2$ is replaced by the sum rule obtained in Ref. \cite{DAs}.

Before arriving at the final numerical values of the parameters from QCD sum rules, we first need to choose the working window of the Borel parameter, which is determined by requiring that both the higher resonance contributions and the higher dimension contributions are subdominant in comparison with the pole contributions. The choice of the Borel mass for different sum rules is presented in Table.\ref{tab:sumrule}. Another important parameter in the sum rules is the threshold, by choosing which the higher resonance contribution can be represented by the integration of the spectral density with the help of quark-hadron duality. The threshold is usually connected with the first resonance having the same quantum number as the concerned composite particle. It is also required that the sum rule does not dependent on the threshold very much. With the above criterion, in the analysis we use $2.65\;\mbox{GeV}^2\leq s_0\leq 2.85\;\mbox{GeV}^2$. Finally, the inputs of the vacuum condensates we used are the standard values: $a=-(2\pi)^2\langle\bar
uu\rangle=0.55\; \mbox{GeV}^{3}$, $b=(2\pi)^2\langle\alpha_sG^2/\pi\rangle=0.47\; \mbox{GeV}^{4}$, $a_s=-(2\pi)^2\langle\bar ss\rangle=0.8a$, $\langle\bar
ug_c\sigma\cdot Gu\rangle=m_0^2\langle\bar uu\rangle$, and $m_0^2=0.8\; \mbox{GeV}^{2}$. The mass of the strange quark is used as $m_s=0.15\,\mbox{GeV}$.
In consideration of the isospin symmetry, the $\Sigma$ mass is used the central value of $\Sigma^+$ presented by the Particle Data Group (PDG) \cite{PDG}: $M_{\Sigma^+}=1.189\mbox{GeV}$. The final results for the nonperturbative parameters are listed in Table.\ref{tab:sumrule}.

\begin{table}
\renewcommand{\arraystretch}{1.1}
\caption{Results from QCD sum rules of the nonperturbative parameters.}
\begin{center}
\begin{tabular}{|l|l|l|l|l|}
\hline
Parameter & $V_1^s$ & $A_1^u$ & $f_1^s$ & $f_2^s$ \\
\hline
$M_B^2(GeV^2)$ & $0.8\sim1.5$ & $1\sim2$ &$0.7\sim0.9$ & $0.7\sim0.9$ \\
\hline
Results & $0.39\pm0.01$ & $0.29\pm0.12$ &$-0.15\pm0.12$ & $9.9\pm2.5$  \\
\hline
Parameter & $f_1^u$ & $f_3^s$ & ${\mathcal P}_2^0(GeV^2)$ & ${\mathcal S}_1^u(GeV^2)$\\
\hline
$M_B^2(GeV^2)$  & $0.7\sim0.9$ & $0.7\sim0.9$ & $0.7\sim0.9$ &$0.7\sim0.9$\\
\hline
Results &  $-0.11\pm0.01$ & $1.6\pm0.2$ & $0.0040\pm0.0004$ &$-0.0014\pm0.0002$ \\
\hline
\end{tabular}
\end{center} \label{tab:sumrule}
\end{table}

\section{Explicit expressions of the $\Sigma$ LCDAs}\label{sec:result}

\begin{figure}
\begin{minipage}{8cm}
\epsfxsize=7cm \centerline{\epsffile{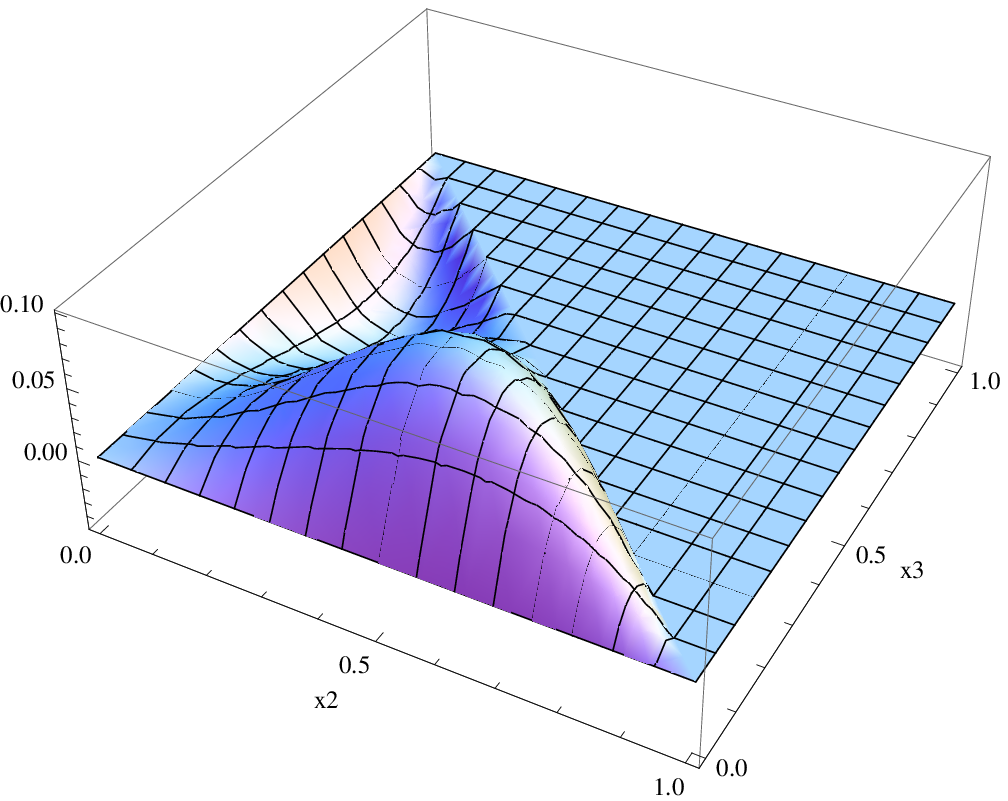}}
\end{minipage}
\begin{minipage}{8cm}
\epsfxsize=7cm \centerline{\epsffile{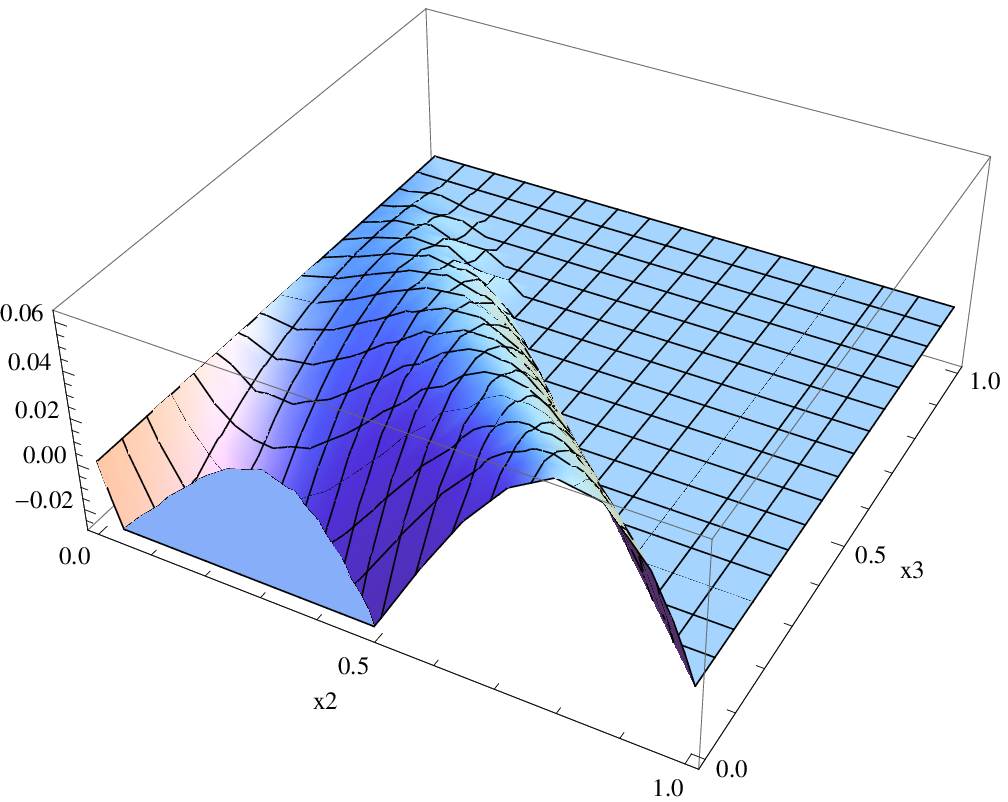}}
\end{minipage}
\caption{\quad Twist-$3$ distribution amplitudes $\Phi_3(x_i)$ (left) and Twist-$4$ distribution amplitudes $\Phi_4(x_i)$ (right).} \label{phi-figure}
\end{figure}

In this section we present the explicit expressions of the $\Sigma$ baryon LCDAs. By considering expressions defined in (\ref{contwist3}) to (\ref{contwist6}), we first plot one of the twist-$3$ distribution amplitude $\Phi_3(x_i)$ and one of the twist-$4$ distribution amplitude $\Phi_4(x_3)$ in Fig. \ref{phi-figure} as an example.

For the definition in (\ref{da-deftwist}), our results are listed as follows: Twist-$3$ distribution amplitudes of $\Sigma$ are
\begin{eqnarray}
V_1(x_i)&=&120x_1x_2x_3[\phi_3^0+\phi_3^+(1-3x_3)],\hspace{1.0cm}A_1(x_i)=-120x_1x_2x_3(x_1-x_2)\phi_3^-,\nonumber\\
T_1(x_i)&=&120x_1x_2x_3[t_1^0+t_1^-(x_1-x_2)+t_1^+(1-3x_3)].
\end{eqnarray}
Twist-$4$ distribution amplitudes are
\begin{eqnarray}
S_1(x_i)&=&6(x_2-x_1)x_3(\xi_4^0+\xi_4^{'0}+\xi_4^++\xi_4^{'+})+6(x_2^2-x_1^2)x_3(\xi_4^-+\xi_4^{'-})\nonumber\\
&&-6(x_2-x_1)x_3^2(\xi_4^-+\xi_4^{'-}),\nonumber\\
P_1(x_i)&=&6(x_2-x_1)x_3(\xi_4^0-\xi_4^{'0}+\xi_4^+-\xi_4^{'+})+6(x_2^2-x_1^2)x_3(\xi_4^--\xi_4^{'-})\nonumber\\
&&-6(x_2-x_1)x_3^2(\xi_4^--\xi_4^{'-}),\nonumber\\
V_2(x_i)&=&24x_1x_2[\phi_4^0+\phi_4^+(1-5x_3)],\hspace{2cm}A_2(x_i)=-24x_1x_2(x_1-x_2)\phi_4^-,\nonumber\\
V_3(x_i)&=&12x_3(1-x_3)[\psi_4^0+\psi_4^+]+12[(x_1^2+x_2^2)x_3-(x_1+x_2)x_3^2]\psi_4^--120x_1x_2x_3\psi_4^+,\nonumber\\
A_3(x_i)&=&-12x_3(x_1-x_2)[\psi_4^0+\psi_4^-]-12(x_1^2-x_2^2)x_3\psi_4^-+12(x_1-x_2)x_3^2\psi_4^-,\nonumber\\
T_2(x_i)&=&24x_1x_2[t_2^0+t_2^-(x_1-x_2)+t_2^+(1-5x_3)],\nonumber\\
T_3(x_i)&=&6x_3(1-x_3)(\xi_4^0+\xi_4^{'0}+\xi_4^++\xi_4^{'+})+6(x_1^2+x_2^2)x_3(\xi_4^-+\xi_4^{'-})\nonumber\\
&&-6(x_1+x_2)x_3^2(\xi_4^-+\xi_4^{'-})-60x_1x_2x_3(\xi_4^++\xi_4^{'+}),\nonumber\\
T_7(x_i)&=&6x_3(1-x_3)(-\xi_4^0+\xi_4^{'0}-\xi_4^++\xi_4^{'+})+6(x_1^2+x_2^2)x_3(-\xi_4^-+\xi_4^{'-})\nonumber\\
&&-6(x_1+x_2)x_3^2(-\xi_4^-+\xi_4^{'-})-60x_1x_2x_3(-\xi_4^++\xi_4^{'+}).
\end{eqnarray}
Twist-$5$ distribution amplitudes are
\begin{eqnarray}
S_2(x_i)&=&\frac32(x_1-x_2)(\xi_5^0+\xi_5^{'0}+\xi_5^++\xi_5^{'+})-3(x_1^2-x_2^2)(\xi_5^++\xi_5^{'+})\nonumber\\
&&-\frac32(x_1-x_2)x_3(\xi_5^-+\xi_5^{'-}),\nonumber\\
P_2(x_i)&=&\frac32(x_1-x_2)(\xi_5^0-\xi_5^{'0}+\xi_5^+-\xi_5^{'+})-3(x_1^2-x_2^2)(\xi_5^+-\xi_5^{'+})\nonumber\\
&&-\frac32(x_1-x_2)x_3(\xi_5^--\xi_5^{'-}),\nonumber\\
V_4(x_i)&=&3(1-x_3)[\psi_5^0+\psi_5^+]+6x_1x_2\psi_5^--3(1-x_3)x_3\psi_5^--6(x_1^2+x_2^2)\psi_5^+,\nonumber\\
A_4(x_i)&=&3(x_1-x_2)[\psi_5^0+\psi_5^+]-3(x_1-x_2)x_3\psi_5^-+6(x_2^2-x_1^2)\psi_5^+,\nonumber\\
V_5(x_i)&=&6x_3[\phi_5^0+\phi_5^+(1-2x_3)],\hspace{2cm}A_5(x_i)=-6x_3(x_1-x_2)\phi_5^-,\nonumber\\
T_4(x_i)&=&-\frac32(x_1+x_2)(\xi_5^{'0}+\xi_5^0+\xi_5^{'+}+\xi_5^+)-\frac32x_1x_2(\xi_5^{'-}+\xi_5^-)\nonumber\\
&&+\frac32(1-x_3)x_3(\xi_5^{'-}+\xi_5^-)+3(x_1^2+x_2^2)(\xi_5^{'+}+\xi_5^+),\nonumber\\
T_5(x_i)&=&6x_3[t_5^0+t_5^-(x_1-x_2)+t_5^+(1-2x_3)],\nonumber\\
T_8(x_i)&=&\frac32(x_1+x_2)(\xi_5^{'0}-\xi_5^0+\xi_5^{'+}-\xi_5^+)+\frac32x_1x_2(\xi_5^{'-}+\xi_5^-)\nonumber\\
&&+\frac32(1-x_3)x_3(\xi_5^{'-}-\xi_5^-)-3(x_1^2+x_2^2)(\xi_5^{'+}-\xi_5^+).
\end{eqnarray}
Finally twist-$6$ distribution amplitudes are
\begin{eqnarray}
V_6(x_i)&=&2[\phi_6^0+\phi_6^+(1-3x_3)],\hspace{2.5cm}A_6(x_i)=-2\phi_6^-(x_1-x_2),\nonumber\\
T_6(x_i)&=&2[t_6^0+t_6^-(x_1-x_2)+t_6^+(1-3x_3)].
\end{eqnarray}

\section{Summary}\label{sec:summary}

The main aim of this work is to present the explicit expressions of the $\Sigma$ baryon light-cone distribution amplitudes. The LCDAs are examined up to twist-$6$ based on the conformal symmetry of the massless QCD Lagrangian. The previous papers indicate that higher conformal spin expansion may contribute in some dynamical processes. Therefore we have to deal with more nonperturbative parameters both at leading order and at next-to-leading order to give more detailed information of the LCDAs of the baryon.

Although we can give a general definition of the LCDAs according to the Lorentz structure of the nonlocal three-quark matrix element between vacuum and the baryon state, we first need to define the independent distribution amplitudes in a proper frame in order to describe them with nonperturbative parameters of QCD. With the help of the conformal symmetry, the LCDAs are redefined and expanded with the conformal spin to the next-to-leading (NL) order in terms of quark fields with definite chirality. In comparison with the case of the nucleon, the number of the independent distribution functions of $\Sigma$ is $14$, which come from the identity symmetry of the two $u$ or $d$ quarks. The NL corrections of the LCDAs come from the next-to-leading order expansion of the nonlocal three-quark operator matrix element. The matrix element is parametrized to the nonperturbative inputs which are connected with the intrinsic properties of QCD. In the calculations, the required nonperturbative inputs are determined in the QCD sum rule approach. We finally present the explicit expressions of the light-cone distribution amplitudes of the $\Sigma$ baryon up to twist $6$ as the main results of this paper.

\acknowledgments  This work was supported in part by the National
Natural Science Foundation of China under Contracts No.11105222 and No.11275268. One of the authors Y. L. Liu also thanks the NSFC program (No.11391240184) and the International Centra of Theoretical Physics (ICTP) for financing the attendance of the summer school sm2463 and sm2466 held in Italy.
\appendix
\section{Equation of motion}\label{app:1}

There are altogether $24$ coefficients when parametrizing the matrix element of the nonlocal three-quark operator. We wish to reduce the number of the independent parameters as far as possible. Fortunately there are relations from the equation of motion of the matrix elements of some different local composite operators. The same relations can be found in Ref. \cite{Braun1}. In this paper we present them only for the completeness of the article and give the direct results from the constraints of these equations. The constraints are:
\begin{eqnarray}
&&
\langle{0}| \epsilon^{ijk} u^i(0) C \gamma_\rho u^j(0)\gamma^\lambda [iD_\lambda s_\gamma]^k(0)|{\Sigma,P} \rangle = 0 \,,
\nonumber \\
&&
\langle{0}| \epsilon^{ijk} u^i(0) C \gamma^\lambda u^j(0)
[iD_\lambda s_\gamma]^k(0)|{\Sigma,P}\rangle =
P_\lambda
\langle{0}| \epsilon^{ijk} u^i(0) C \gamma_\lambda u^j(0)
s_\gamma^k(0)|{\Sigma,P}\rangle \,,
\nonumber \\
&&
\langle{0}| \epsilon^{ijk} u^i(0) C \sigma_{\alpha\beta}  u^j(0)
\gamma^\lambda [iD_\lambda s_\gamma]^k(0)|{\Sigma,P}\rangle = 0 \,,
\nonumber \\ &&
\langle{0}| \epsilon^{ijk} u^i(0) C i \sigma_{\alpha\beta}  u^j(0)
[iD^\beta s_\gamma]^k(0)|{\Sigma,P}\rangle \nonumber \\ &&
= P^\beta \langle{0} \epsilon^{ijk} u^i(0) C i \sigma_{\alpha\beta}  u^j(0)s^k(0) |{\Sigma,P}\rangle
-\langle{0} \epsilon^{ijk} [u(0) C i\stackrel{\leftrightarrow}{D}_\alpha  u(0)]^{ij} s_\gamma^k(0) |{\Sigma,P}\rangle \,,
\nonumber \\ &&
\langle{0}| \epsilon^{ijk} u^i(0) C i \gamma_5\sigma_{\alpha\beta}  u^j(0) [iD^\beta s_\gamma]^k(0)|{\Sigma,P}\rangle \nonumber \\ &&
= P^\beta \langle{0} \epsilon^{ijk} u^i(0) C \gamma_5 i \sigma_{\alpha\beta} u^j(0)
s^k(0)|{\Sigma,P}\rangle-\langle{0}| \epsilon^{ijk} [u(0) C \gamma_5 i \stackrel{\leftrightarrow}{D}_\alpha  u(0) ]^{ij}s_\gamma^k(0)| {\Sigma,P}|\rangle\,,
\nonumber \\
&&
\langle{0}| \epsilon^{ijk} [u(0) C \gamma^\rho \gamma_5 \stackrel{\leftrightarrow}{D}_\rho u(0) ]^{ij}d_\gamma^k(0)|{\Sigma,P}\rangle = 0
\,,
\nonumber \\
&&\langle{0}| \epsilon^{ijk} [u(0) C \{\gamma_\lambda i\stackrel{\leftrightarrow}{D}_\rho -
\gamma_\rho i\stackrel{\leftrightarrow}{D}_\lambda\} \gamma_5 u(0)]^{ij}
s_\gamma^k(0)| {\Sigma,P}\rangle \nonumber
\\ &&
=
\langle{0}| \epsilon^{ijk} [u(0) C
\frac{i}{2}\{\sigma_{\lambda\rho} \gamma^\alpha i\overleftarrow{D}_\alpha +
\gamma^\alpha \sigma_{\lambda\rho} i\overrightarrow{D}_\alpha\} \gamma_5 u(0)]^{ij}s_\gamma^k(0)| {\Sigma,P} \nonumber\\ &&
=- i \epsilon_{\lambda\rho\alpha\delta} [P^\alpha
\langle{0}| \epsilon^{ijk} u^i(0) C \gamma^\delta u^j(0) d_\gamma^k(0)| {\Sigma,P}\rangle
-\langle{0}| \epsilon^{ijk} u^i(0) C \gamma^\delta u^j(0)
[i D^\alpha s_\gamma]^k(0)| {\Sigma,P}] \,.
\nonumber \\
\end{eqnarray}

A simple calculation leads to the following relationships:
\begin{eqnarray}
&\mathcal{V}_1^s=4\mathcal{V}_2^0+2\mathcal{V}_3^s,\, &-3\mathcal{V}_5^0=3\mathcal{V}_3^s+\mathcal{V}_4^0,\nonumber\\
&\mathcal{V}_1^0-\mathcal{V}_3^0=\mathcal{V}_1^s-\mathcal{V}_3^s+4\mathcal{V}_4^0-\mathcal{V}_2^0,\, &\mathcal{T}_3^s+\mathcal{T}_5^0=0,\nonumber\\
&\mathcal{T}_2^0=-\mathcal{T}_1^s+4\mathcal{T}_3^s+3\mathcal{T}_4^0,\, &\mathcal{T}_1^0-2\mathcal{T}_3^0-\mathcal{S}_2^0=\mathcal{T}_1^s
-2\mathcal{T}_3^s-\mathcal{T}_4^0+3\mathcal{T}_5^0+6\mathcal{T}_7^0\nonumber\\
&\mathcal{T}_1^0-2\mathcal{T}_3^0-\mathcal{S}_1^u=\mathcal{T}_1^s
-3\mathcal{T}_2^0-2\mathcal{T}_3^s-\mathcal{T}_4^0,\, &
-2\mathcal{T}_3^s+2\mathcal{T}_4^0=-2\mathcal{T}_3^0-\mathcal{P}_1^u,\nonumber\\
&2\mathcal{T}_3^0-\mathcal{P}_2^0=2\mathcal{T}_3^s-2\mathcal{T}_4^0-6\mathcal{T}_7^0\, &
\mathcal{A}_1^u+\mathcal{A}_3^u+\mathcal{A}_2^0+4\mathcal{A}_4^0=0,\nonumber\\
&\mathcal{A}_3^u-\mathcal{A}_2^0=\mathcal{V}_2^0+\mathcal{V}_3^0-\mathcal{V}_3^s,\, &
2\mathcal{A}_5^0=\mathcal{V}_2^0+\mathcal{V}_3^0-2\mathcal{V}_5^0-\mathcal{V}_3^s.\label{eom}
\end{eqnarray}

\section{QCD sum rules of the nonperturbative parameters}\label{app:2}
In this Appendix we introduce the QCD sum rule method of the nonperturbative parameters which are required in the paper. We take the process for the decay constant $V_1^s$ as an example. It starts from the following correlation function:
\begin{equation}
\Pi(q)=i\int d^4xe^{iq\cdot z}\langle|0j_{1s}(x)\bar j_1(0)|\rangle,
\end{equation}
where $j_{1s}(x)=\epsilon^{ijk}[u^i(x)C\!\not\!{z}u^j(x)]\gamma_5[iz\overrightarrow Ds(x)]^k$, and $j_1(0)=\epsilon^{ijk}[u^i(0)C\!\not\!{z}u^j(x)]\gamma_5s^k(0)$.
In compliance with the general process of the QCD sum rules, we need to calculate the correlation function both phenomenally and theoretically. On the phenomenon side, we interpolate a complete set of states with the same quantum number as the $\Sigma$ baryon to get the hadronic representation
\begin{equation}
\Pi(q)=\frac{2f_{\Sigma}^2V_1^s(q\cdot z)^4\!\not\!{z}}{M^2-q^2}+...,
\end{equation}
where ``..." represents contribution from higher resonances and continuum states.
By making use of the dispersion relation, the equation above can be written as the integration form,
\begin{equation}
\Pi(q)=\frac{2f_{\Sigma}^2V_1^s(q\cdot z)^4\!\not\!{z}}{M^2-q^2}+\int_{s_0}^\infty \frac{\frac{1}{\pi}Im \Pi(s)}{s-q^2}ds.\label{disper}
\end{equation}

On the theoretical side, we calculate the correlation function at the quark level by use of the operator product expansion (OPE). In the calculation we expand up to dimension $6$ accuracy. Then by hadron-quark duality approximation, the integration function in (\ref{disper}) can be equalized by the spectral density calculated theoretically.

As the two representations have the same content, they can be matched so as to get the sum rule. Additionally, in order to make the numerical estimation more accurate, we use the Borel transformation to suppress both higher resonances and higher dimensional contributions. The Borel transformation is defined as
\begin{equation}
\hat{B}^{Q^2}_{M_B^2}\equiv \lim_{Q^2\rightarrow \infty, N\rightarrow \infty}(-Q^2)^N(\frac{d}{dQ^2})^N.
\end{equation}

\begin{figure}
\begin{minipage}{8cm}
\epsfxsize=7cm \centerline{\epsffile{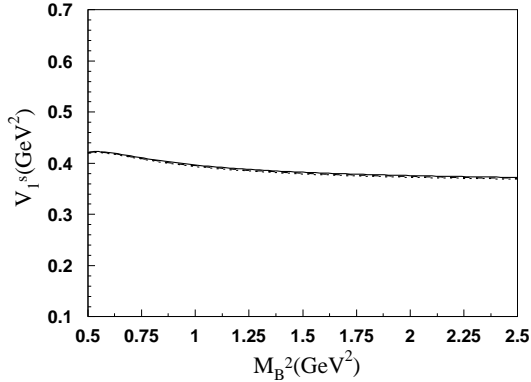}}
\end{minipage}
\caption{\quad Borel working window $V_1^s$ with threshold $2.65\;\mbox{GeV}^2\leq s_0\leq 2.85\;\mbox{GeV}^2$ from up down.} \label{v1s}
\end{figure}

Before getting the numerical estimates of the hadronic parameter, we still have to determine the necessary input parameters such as the threshold $s_0$ and the Borel mass $M_B^2$. The threshold is the point from which the higher resonance contributions can be described by the integration of the spectral density, so it is connected with the first resonance state that has the same quantum number as the hadron we concern. In the numerical analysis, we use the values $s_0=(2.65 \sim 2.85)\;\mbox{GeV}^{2}$. The Borel mass is determined by two different requirements. First, it should be large enough so that the higher dimension contributions can be suppressed efficiently, namely, the OPE approach is satisfied. Therefore, we give the down limit of Borel mass by requiring the higher dimension contributions is less than $ 30\% $ of the whole. Second, the Borel mass needs to be small so that the higher resonance contributions can be suppressed efficiently. In determining the up limit of the parameter it is required that the pole contribution is larger than that of the higher resonances. In Fig. \ref{v1s} we plot the numerical results with the Borel window for the sum rule of $V_1^s$. It is shown that in the working window $0.8\mbox{GeV}^{2}\leq M_B^2\leq 1.4\;\mbox{GeV}^{2}$ the sum rule is acceptable, so that we arrive at the estimate of the coupling constant
\begin{equation}
V_1^s=0.39\pm0.01,
\end{equation}
where the error comes from the uncertainties of both the Borel parameter and the threshold $s_0$. The other sum rules obtained in the previous text can be analyzed in the similar processes and the numerical results are illustrated in Table. \ref{tab:sumrule}.

\end{document}